\documentclass{article}

\usepackage{arxiv}

\usepackage[utf8]{inputenc} 
\usepackage[T1]{fontenc}    
\usepackage{hyperref}       
\usepackage{url}            
\usepackage{booktabs}       
\usepackage{amsfonts}       
\usepackage{nicefrac}       
\usepackage{microtype}      
\usepackage{lipsum}
\usepackage{float}
\usepackage{listings}
\usepackage{graphicx}
\usepackage{balance} 

\title{SAVIME: A Multidimensional System for the Analysis and Visualization of Simulation Data}

\author{
  Hermano L. S. Lustosa \\
  DEXL Laboratory\\
  National Laboratory for Scientific Computing\\
  Petrópolis, RJ - Brazil \\
  \texttt{hermano@lncc.br} \\
   \And
  Fabio Porto\\
  DEXL Laboratory\\
  National Laboratory for Scientific Computing\\
  Petrópolis, RJ - Brazil \\
  \texttt{fporto@lncc.br} \\
}

\begin{document}
\maketitle

\begin{abstract}
Scientific applications produce a huge amount of data, which imposes serious management and analysis challenges. In particular, limitations in current database management systems prevent their adoption in simulation applications, in which in-situ analysis libraries, in-transit I/O interfaces and scientific format files are preferred over DBMSs. In order to make simulation applications benefit from DBMS support, the author proposes the development of a system called SAVIME in the context of his PhD thesis. SAVIME is an array database system designed to manage numerical simulation data. In this document, the author presents all work conducted so far and the current state of development.
\end{abstract}


\section{Introduction}

The ever increasing computational power of HPC environments allows for performing very complex numerical simulations for many phenomena. These simulations produce huge datasets that need to be analyzed and visualized to enable researchers to gain insights about the phenomena being studied.

Traditionally, researchers adopt the post-processing approach for analysis and visualization creation of simulation data. The simulation code stores its raw data in a file system, and another application reads it from disk, performs analysis and creates the visualization files. This is the easiest possible approach, which offers the best separation between the simulation code and the rest of the workflow. However, due to the I/O gap in HPC environments (the fact that storage technologies are improved in a slower rate than processors), storing data on disk and reading it back is deemed as very inefficient for large scale simulations \cite{7064669}. 

Two popular approaches, in-situ and in-transit analysis \cite{Oldfield:2014:EMI:2597652.2597668}, have been proposed to address this problem by avoiding executing costly I/O operations. In-situ analysis consists of adding post-processing activities directly to the simulation code, thus, allowing analysis to be executed on data in memory. This approach does not require the entire simulation dataset to be written to and read from disk, but brings other challenges. The same computational resources used for running the simulation is available for activities executed during analysis and visualization, and balancing the use of these resources in a manner that the highly tuned simulation code is not severely affected requires a lot of work. 

In-transit analysis is less invasive. In this approach, a different partition of the computational resources is used to execute analysis just like in post-processing, but instead of relying on the file system for communication, processes use I/O libraries and high-speed network facilities. Even though it yields less interference with simulation code, in-transit analysis requires some method to coordinate the allocation of additional resources during simulation execution.

An important point to consider is that DBMSs are not commonly adopted in any of these approaches. In-situ analysis relies on special libraries, such as Paraview Catalyst \cite{Ayachit:2015:PCE:2828612.2828624}. In-transit analysis requires the use of I/O interfaces, such as ADIOS \cite{5161052} and GLEAN \cite{Vishwanath:2011:TDM:2063384.2063409}. The post-processing approach consists of storing data in scientific data formats such as HDF\cite{hdf5} and NetCDF\cite{netcdf}.

DBMSs are considered inefficient for scientific data management due to the impedance mismatch problem \cite{Blanas:2014:PDA:2588555.2612185, Gosink:2006:HAC:1154779.1154994}, i.e., the incompatibilities between the representation formats of the source data and the DBMS. This impedance mismatch yields costly conversions between formats, which adds prohibitive overhead during data analysis. Therefore, scientific file formats are preferred over DBMSs for maintaining and analyzing scientific datasets, even though these file formats do not offer the analytic capabilities found in a DBMS.

Historically, DBMSs have offered many advantages over simply maintaining data in files, e.g., DBMSs offer declarative query languages, which eases data analysis and avoids the need of extensive coding/scripting, and a logical data view that isolates data from the applications that consume it. Furthermore, given that an appropriate data model is in place, a DBMS provides users with a semantically richer data representation, which simplifies the creation of analytical queries and visualizations. 

In \cite{Lustosa:2016:DSS:3007263.3007271}, many difficulties regarding simulation data representation were identified. The conclusion is that current DBMSs solutions, even the ones designed for scientific tasks, implement a very restrictive data model no well suited for simulation data. In addition, to perform visualization (which is paramount for understanding the results) with data stored in current DBMS solutions, it is necessary to retrieve the attributes of interest, copying large datasets between memory spaces and carrying out another costly data conversion for the output visualization format. 

Given the current lack of a database solution that could facilitate the simulation data management, analysis and visualization, the author proposes the research and development of an extension of the array data model \cite{lustosa:lirmm-01620376} named TARS, to cope with simulation data features and to allow a more efficient representation, along with a system that implements this model, currently named 
SAVIME. 

Therefore, the research goals for this thesis (and consequently for the system SAVIME) involve the identification and/or development and implementation of techniques in order to fulfill the following requisites:

\begin{enumerate}
 \item Fast data ingestion without impedance mismatch, i.e., the ability to cope with data as it is generated by the simulation code without carrying out costly data conversions during data ingestion.
 \item Design of a data model which not only offers an elegant representation for simulation data, but also allows SAVIME to take advantage of the preexisting data partitioning and sorting in order to process queries efficiently.
 \item Integration with in-situ libraries, which enables the analysis and visualization code to harvest data directly from SAVIME's memory space, thus, avoiding the overhead of creating a separate application to query the DBMS and convert query results to the visualization file format.
\end{enumerate}

At the current state of the development, the author already has the foundations of the system SAVIME implemented and evaluated. Therefore, this document will present the first achievements obtained, as well as the following steps to be taken for the thesis completion.

This document is organized as follows. Section 2 presents the related work. Section 3 presents the system SAVIME. Section 4 presents an explanation for a visualization use case. Section 5 contains the preliminary experimental evaluation for the current state of development. Finally, Section 6 concludes.

\section{Related Work}

\subsection{Array Databases}

The definition of the first array data models and query languages dates back to the works of Baumann \cite{Baumann:1994:MMD:615204.615207} and Marathe \cite{Marathe:1997:LMA:645923.671002} \cite{marathe1999query}. Since that time, a myriad of systems emerged in order to allow for the storage and analysis of data over multidimensional arrays, among them are:

\begin{itemize}
 \item Titan \cite{581883} is a parallel DBMS designed to deal with remote sensor data obtained by satellites. It has a standard architecture for parallel DBMSs, with a single node acting as a coordinator node, and the other nodes functioning as workers. The coordinator node receives the queries and schedules their processing in the worker nodes.
 
 \item RasDaMan \cite{Baumann:1997:RAM:331697.331732} is DBMS which supports arrays with an arbitrary number of dimensions. RasDaMan works as a middleware over a relational DBMS, implementing a common client-server architecture in a single machine.
 
 \item ArrayDB \cite{marathe1999query} is a prototype array DBMS that implements the AML language, based on UDFs (user defined functions). The execution of an AML query is based on a pipeline in which data moves through the operators piece by piece. The order in which these operators are applied over data is defined by the query optimizer, a software component that defines the most efficient query plan to be executed.
 
 \item (S)RAM \cite{Cornacchia:2008:FEI:1325148.1325166} is a DBMS that implements an array based algebra over MonetDB. Therefore, the operations over matrices are mapped to relational algebra operations, like projections, joins, groupings among others.
 
 \item SciDB \cite{Cudre-Mauroux:2009:DSS:1687553.1687584} represents the state of the art in systems for managing dense multidimensional arrays (even though it also supports sparse matrices) with an arbitrary number of dimensions. SciDB has its AFL and AQL query languages, is natively distributed and supports most common array operations out of the box.
 
\end{itemize}

Even though many array systems exist, their array data models have limitations that hamper the representation of scientific datasets, like those originated by simulations\cite{lustosa:lirmm-01620376}. Thus, the need for a system like SAVIME, as proposed in the author's PhD thesis.

\subsection{Scientific Formats, Libraries and Systems}

Another issue regarding DBMS usage for scientific data management is related to the impedance mismatch problem. The underlying data models and storage schemes of a DBMS might differ greatly from the source format the data is initially generated, which forces users to convert their datasets to a compatible representation in order to load them into the system. This problem is common in scientific application domains. For instance, loading and indexing simulation datasets into a DBMS can incur high overhead, thus preventing scientists from adopting any DBMS at all. Instead, scientists typically rely on I/O libraries that provide support for multidimensional arrays, e.g., HDF\cite{hdf5} and NetCDF\cite{netcdf}. These libraries give users more control over their data without incurring the performance penalties for data loading in a DBMS \cite{Blanas:2014:PDA:2588555.2612185, Gosink:2006:HAC:1154779.1154994}. They are also flexible, allowing users to specify how their data (produced by simulations) is laid out and avoid the expensive conversions performed by the DBMS during data ingestion.

However, I/O libraries do not offer all benefits that a full-fledged DBMS does. These benefits include, but are not limited to features, such as query languages, data views and the isolation between data and applications that consume data. NoDB \cite{Alagiannis:2012:NEQ:2213836.2213864} is a first attempt to bridge the gap between DBMS high ingestion costs and I/O libraries access efficiency for relational DBMSs. The approach advocates that the DBMS should be able to work with data as laid out by the data producer, with no overhead for data ingestion and indexing. Any subsequent data transformation or indexing performed by the DBMS in order to improve the performance of data analyses should be done adaptively as queries are submitted. We believe that even though NoDB is currently implemented on top of a RDBMS and lacks support for multidimensional data, its philosophy can be successfully applied for array databases as well. In this context, SAVIME provides means for fast data ingestion free from costly data conversions common to other array DBMSs. The TARS data model allows huge memory chunks that contain arrays of numerical values output from solvers of linear systems to be efficiently ingested into the system without any type of rearrangements. 

The challenges of monitoring, debugging, steering and analyzing scientific applications dataflows have been tackled by DfAnalyzer \cite{dfanalyzer}. DfAnalyzer allows for combining provenance data with data generated by the scientific dataflow providing means for executing a myriad of important analysis. In this regard, SAVIME is a system focused solely on representing, storing and analyzing data efficiently, without necessarily the concern of capturing provenance data and monitoring execution, which is a much broader problem.

From a data model perspective, the challenges of representing simulation data efficiently have been tackled in many works \cite{RezaeiMahdiraji:2013:IGD:2541167.2505733} \cite{Howe06gridfields:model-driven} \cite{Lee99i.l.:modeling}. Some particular data models have been proposed, but none of them was established as a standard for simulation data. In general, arrays are considered the most common format for scientific data. Scientific databases and the aforementioned I/O libraries support arrays. Even scientific visualization libraries, such as VTK \cite{vtk} that enables the most varied grids to be represented and visualized, has arrays (called VTKArrays) as its underlying data structure.

In SAVIME, the idiosyncrasies of simulation data are captured through the usage of specially defined arrays with associated semantical annotations. These  or typed arrays conform with a defined type that represents some aspect of grid data, such as geometries topologies or field data.

\subsection{In-situ, In-transit and Pos-processing Analysis}

Traditionally, numerical simulation analysis and visualization occur in the post processing step of the scientific workflow. The simulation code stores its raw data on the file system, and another application reads it from disk, performs analysis and creates the visualizations. This is the easiest possible approach, offering the best separation between the simulation and the rest of the workflow. However, due to the I/O gap in HPC environments, storing data to disk and reading it back is deemed as inefficient  for large scale simulations.

For regular post-processing, scientists make use of scientific file formats, such as HDF5, NetCDF, etc. The impedance mismatch problem, common in DBMSs, is attenuated with the use of scientific file formats such as HDF5 because it is easier to move large datasets from memory to the files using I/O libraries. HDF5 is a widely accepted format, compliant with many different applications, and it is also enables the use MPI-IO along with a distributed file system to improve performance.

HDF5 format stores data on a tree-like structure whose nodes are mainly groups or datasets. Groups are like directories in a file system, and they can contain a set of datasets or others groups. Datasets are the objects that actually store data as multidimensional arrays. Data from every simulation trial can be stored on a different HDF5 group. Under the group, the raw data can be stored on its corresponding datasets for each existing field data. Unidimensional arrays can be used for non-transient problems and bidimensional arrays for transient problems, adding another dimension for time series values. It is also possible to adjust the schema in HDF5 to represent other data and metadata, such as the input parameters.

Independently from the raw data format, indexing is required to accelerate queries on scientific dataset. Bitmap indexes are considered ideal for scientific workloads that are read-predominant. Past works \cite{99treebased} have predicted that bitmaps would be preferred over tree-based multidimensional indexes for applications that do not require many data changes. More recently \cite{Gosink:2006:HAC:1154779.1154994}, there has been an effort to improve queries on HDF5 with bitmap indexes.

Despite the efforts for querying HDF5 efficiently with indexes, storing huge simulation datasets by itself is not desirable due to the previously cited I/O constraints in HPC environments. Indexes only make the problem worse, since they significantly increase disk footprint. With that in mind, some other works \cite{Lakshminarasimhan:2013:SSS:2493123.2465527} propose smart encoding and indexing of data, that would reduce it to a more manageable size. Alternatively, sampling \cite{10.1111:j.1467-8659.2011.01964.x} simulation data before storing it might be the more efficient solution.

In order to avoid costly I/O operations, scientists can do in-situ analysis and visualization. In this scenario, it is necessary to add post-processing activities directly to the simulation code, thus, allowing analysis to be executed with data in memory. This method does not require the entire simulation dataset to be written and read from disk, but poses other challenges. The same computational resources used for running the simulation is available for activities executed only on the post-processing step, and balancing the use of these resources in a manner that the highly tuned simulation code is not severely affected requires some work.

In-situ visualization libraries, such as Libsim and Catalyst, allow users to couple their simulation code directly to a visualization tool. Libsim \cite{EGPGV:EGPGV11:101-109} is library that allows a fully-featured visualization tool, VisIt, to request data as needed from the simulation and apply visualization algorithms in situ with minimal modification to the application code.

For coupling simulation code with Paraview, there is another library. Paraview Catalyst \cite{Ayachit:2015:PCE:2828612.2828624} is a data processing and visualization library that enables in situ analysis and visualization. Built on and designed to interoperate with the standard visualization toolkit VTK and the ParaView application, Catalyst enables simulations to intelligently perform analysis, generate relevant output data, and visualize results concurrent with a running simulation.

In-transit analysis is a less invasive procedure. In this scenario, a different partition of the computational resources is used to execute analysis just like in post-processing, but instead of relying on the file system for communication, processes use I/O libraries and high-speed network facilities. Even though it offers less interference with simulation code, in-transit analysis requires some method for coordinate the extra allocation of resources during simulation execution.

In \cite{citeulike:9972483}, several in-transit tools are discussed, such as NeSSIE, GLEAN and ADIOS. The NEtwork Scalable Service Interface (Nessie) \cite{4100403} is a framework for developing in transit analysis capabilities. It provides a remote-procedure call (RPC) abstraction that allows the application-developer to create custom data services to match the specific needs of the application.

GLEAN is a flexible and extensible framework that takes into account application, analysis and system characteristics in order to facilitate simulation-time data analysis and I/O acceleration \cite{Vishwanath:2011:TDM:2063384.2063409}.

The Adaptable I/O \cite{5161052} System framework (ADIOS) is designed to separate the I/O API from the actual implementation of the I/O methods. This design specification enables the users to easily, and without any application source code modifications, select I/O methods that are optimized for performance and functionality on the target platform. These frameworks work as bridges between simulation applications and analysis code. Data is "captured" during I/O and rerouted to the next service in the pipeline completely bypassing disk storage.

The system SAVIME is designed to enable mainly the in-transit approach for data analysis, since it could serve as a staging platform for data. Nevertheless, SAVIME can also perform the in-situ analysis, since it relies in a flexible data model, and in the usage on memory mapped files based on shared-memory, which allows for capturing data in the computation job address space.

\section{The System SAVIME} 

This section presents an in-depth description of the current state of the project. The author presents the TARS data model, and the system SAVIME. In addition, the preliminary experimental evaluations carried out to assess SAVIME's performance are also briefly discussed.

\subsection{Typed Array Data Model} \label{tars}

Scientific data is usually represented as multidimensional arrays. Multidimensional values are a data pattern that emerges in scientific experiments and measurements and can also be generated by simulations. From a data management and representation perspective, the definition of the array data model is presented in \cite{marathe1999query}. In short, an array is a regular structure formed by a list of dimensions. A set of indexes for all dimensions identifies a cell or tuple that contains values for a set of array attributes. 

If carefully designed, arrays offer many advantages when compared to simple bi-dimensional tables. Cells in an array have an implicit order defined by how the array data is laid out in linear storage. We can have row-major, column-major or any other arbitrary dimension ordering. Array DBMSs can quickly lookup data and carry out range queries by taking advantage of this implicit ordering. For instance, suppose that the user wants to extract a slice of an array. Given that data is sorted by its dimension indexes, one can know exactly where the desired data is located without the need for an explicit index structure or worse, a full scan.

If the data follows a well behaved array-like pattern, using arrays saves a lot of storage space, since dense arrays indexes do not need to be explicitly stored. Furthermore, arrays can be split into subarrays, usually called tiles or chunks. These subarrays are used as processing and storage data units. They help answering queries rapidly and enforce a coherent multidimensional data representation in linear storage.

However, current array data model implementations, e.g.,  SciDB and RasDaMan, have some limitations, preventing an efficient representation of simulation datasets. In SciDB \cite{scidb} for instance, it might be necessary to preload multidimensional data into an unidimensional array and then rearrange it during data loading. RasDaMan requires either the creation of a script or the generation of compatible file formats for data ingestion. This may also require costly ASCII to binary conversion (since numerical data is likely to be created in binary format) for adjusting the data to the final representation on disk. In both cases, the amount of work for loading the dataset alone is proportional to its size, making it impractical for the multi-terabyte data generated by modern simulation applications.

Furthermore, the array data model does not explicitly incorporate the existence of dimensions whose indexes are non-integer values. In some simulation applications, the data generated follows an array-like pattern, but one of the identifiable dimensions can be actually a non-integer attribute.  For instance, in 3D rectilinear regular meshes, we have points distributed in spatial dimensions whose indexes or coordinate values are usually floating point numbers. To address this issue, we need to map non-integer values into integer indexes that specify positions within the array. Array DBMSs like SciDB or RasDaMan do not support this kind of functionality currently.

Arrays can be sparse, meaning that there is no data values for every single array cell. Data may also have some variations in their sparsity from a portion of the array to another. This is the case for complex unstructured meshes geometry (with an irregular point distribution in space) when directly mapped to arrays. SciDB provides support to sparse arrays, but since it splits an array into chunks (equally sized subarrays), it is very hard to define a balanced partitioning scheme, because data can be distributed very irregularly. RasDaMan is more flexible in this aspect, and allows arrays to be split into tiles or chunks with variable sizes.
   
Another characteristic of complex multidimensional data representation is the existence of partial functional dependencies with respect to the set of indexes. Partial dependencies occur in constant or varying mesh geometries and topologies, or any other kind of data that does not necessarily varies along all array dimensions. For instance, when researchers create models for simulating transient problems, the time is a relevant dimension to all data, i.e.,  model predictions vary over time. However, the mesh, which is the representation of the spatial domain, may not change in time, meaning that the coordinate values and topology incidence remain the same throughout the entire simulation. Another possibility is the usage of the same mesh for a range of trials, and another mesh for another range. In both cases,  there is a mesh for every single time step (an index in the array time dimension) or trial, but actually only one mesh representation needs to be stored for an entire range of indexes.
  
\subsubsection{Model Description}

The TARS data model extends the basic array data model to cope with complex multidimensional data addressing the issues aforementioned. 

\subsubsection*{TAR Schema (TARS)}

A TAR Schema (TARS) contains a set of Typed ARrays (TAR). A TAR has a set of data elements: dimensions and attributes. A TAR cell is a tuple of attributes accessed by a set of indexes. These indexes define the cell location within the TAR. A TAR has a type, formed by a set of roles. A role in a type defines a special purpose data element with specific semantics. If a TAR is of a given type $T$, it is guaranteed to have a set of data elements that fulfill the roles defined in $T$. This facilitates the creation of operations that require additional semantics about the data. 

\subsubsection*{Mapping Functions}

\begin{figure}[ht]
	\centering
	\includegraphics[width=3.3in]{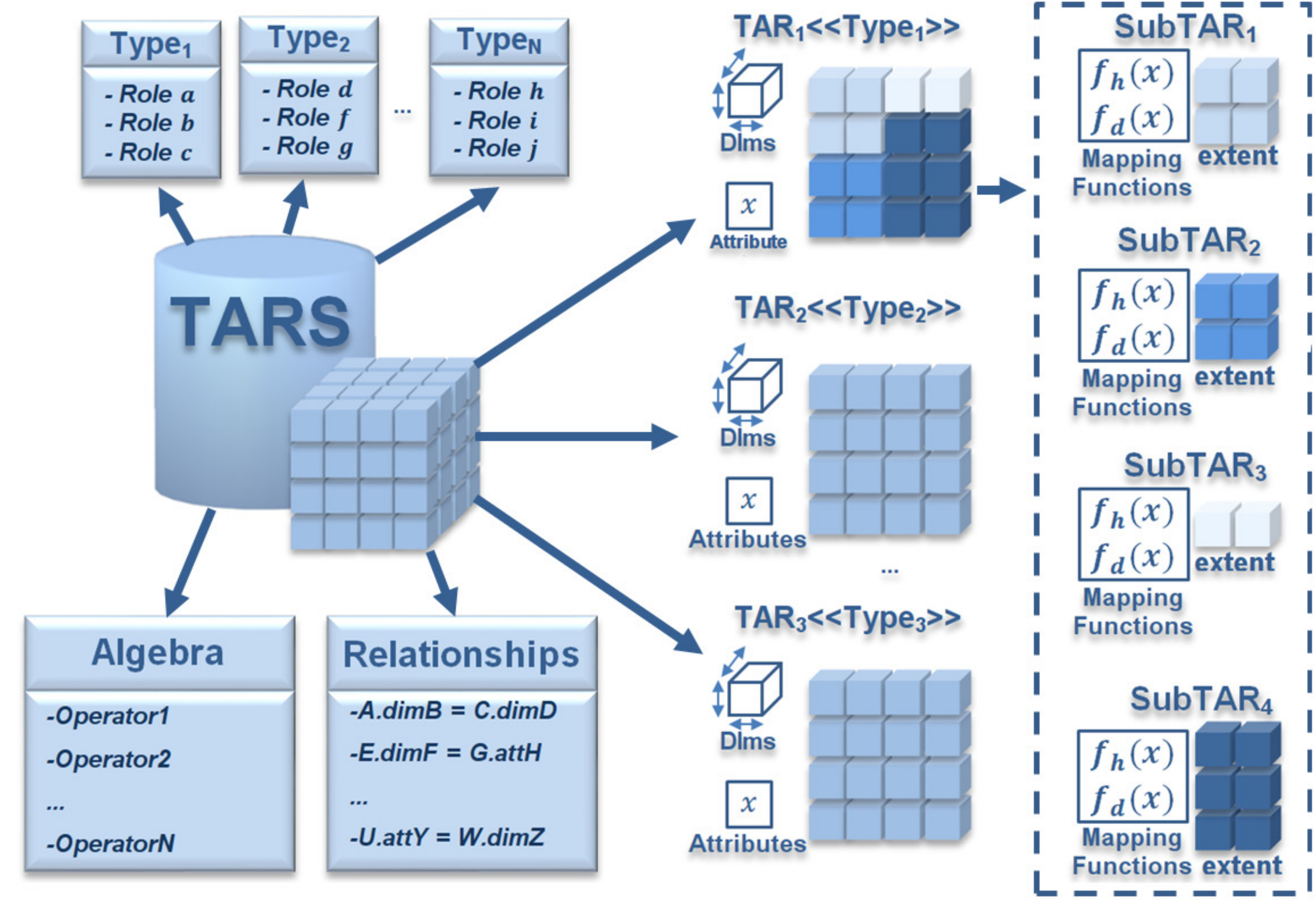}
	\caption{Typed Array Schema Elements}
	\label{fig:tars}
\end{figure}

In TARS, we define mapping functions as a way to provide support for sparse arrays, non-integer dimensions, heterogeneous memory layouts and functional partial dependencies with respect to dimensions. Figure \ref{fig:tars} gives a graphical view of the model.

\subsubsection*{Physical To Logical Representation}

A TAR is split into a set of SubTARs. The latter is said to cover a n-dimensional slice of the TAR, or as we call it in the model, a TAR region. Every subTAR is defined by the TAR region it represents and two mapping functions: position mapping function and data mapping function. 

The position mapping function reflects the actual data layout in memory, since it defines where every TAR cell within a given subTAR ended in linear storage. Therefore, the position mapping function should implement the multidimensional linearization technique used for the data. It can incorporate an n-dimensional to linear translation based on how dimensions are ordered (the case for a row-major or column-major scheme) or even consider a space filling curve approach (e.g. Z-order). 

The data mapping functions translate a linear address into actual data values. In a simple scenario, this function does basically a lookup into a linear array that stores the data. In a more complex scenario, it could compute a derived value from the actual subTAR data. The actual implementation of mapping functions is done in various forms. In some cases, it may be necessary to explicitly store the mapping for every value. In other situations, the mapping occurs in a regular pattern, so translation is done by a single parameterized procedure, in which case only the parameters need to be stored. SubTARs do not only define a partitioning scheme for a TAR, but also serve as a way to allow users to specify the details about how their data is laid out, avoiding costly data transformations and rearrangements during ingestion into the DBMS.

\subsubsection{Formalization}

For completeness, we present here the formalization of the TARS data model. A TARS denoted by $\Gamma$ is a quintuple $(\Omega_{\Gamma}, R_{\Gamma}, T_{\Gamma}, L_{\Gamma}, \Theta_{\Gamma})$ where $\Omega_{\Gamma}$ is a set of typed arrays, $ R_{\Gamma}$ is a set of roles, $T_{\Gamma}$ is a set of types, $L_{\Gamma}$ is a set of links or relationships, and $\Theta_{\Gamma}$ is a set of operators forming a TAR algebra. 

A Typed Array (TAR) $A \in \Omega_{\Gamma}$ is a septuple $(N_A$, $D_A$, $S_A$, $C_A$, $R{_A}$, $\Phi_{A}$, $\Upsilon_{A})$, where $N_A$ is a string containing the TAR name, $D_A$ is a set of data elements, $S_A$ is a set of subTARs, $C_A$ is a set of locations forming the TAR location space, $R{_A}$ is the subTARs location function, $\Phi_{A}$ is the dimension data mapping function, and $\Upsilon_{A}$ is the role mapping function.

A data element $e \in D_A$ is a triple $(N_e, V_e, D_e)$, where $N_e$ is a string value defining the data element name, $V_e$ and $D_e$ are sets of atomic values such that $V_e \subset D_e$. $D_e$ is the domain of the data element, representing the set of all possible values a TAR can hold in that data element, and $V_e$ is the data element image, containing the actual set of values for a data element at any given time. 

$D_A$ can be divided into two subsets, $Dim_A \subset D_A$ containing data elements that are dimensions and $Att_A \subset D_A$ containing data elements that are tuple attributes. We also have that $Dim_A \cap Att_A = \emptyset$, meaning that a data element is either a dimension or an attribute and never both.

A subTAR $s \in S_A$ is a triple $(\eta_s, Fi_s, Fd_s)$ where $\eta_s$ is the set of TAR locations that represents the extent of the subTAR, $Fi_s$ is the subTAR position mapping function, and $Fd_s$ is the subTAR data mapping function. The function $\Phi_{A}$ in the TAR definitions maps a set of dimension values $V_{di}$ for every dimension $d_i \in Dim_A$ to a set of $x_i \in \mathbb{Z}$:

\begin{equation}
\Phi_{A}: (V_{d1} \times V_{d2} \times ... \times V_{dn}) \rightarrow \mathbb{Z}^n
\end{equation}

A location $L_A$ is a position in the $A$ TAR defined as a set of integers coordinate values obtained through the application of $\Phi_{A}$ in a set of dimension values. An array location is nothing more than a multidimensional address formed by a set integer indexes that identifies a tuple or cell within the TAR. The definition of $\Phi_{A}$ is trivial when ${d_i}_{D_i} = \mathbb{Z} \ \forall \ d_i \in Dim_A$.
Every value held in a TAR $A$ can be specified by a location in $A$ and the specification of the data element (either a dimension or attribute). The location space $C_A$ is a set formed by all possible locations in an array given by the image of the function $\Phi_{A}$. 

A subTAR $s \in S_A$ holds the functions to translate a location in a TAR into an address in a linear storage scheme, and then to translate the linear address into a data value valid for an extent of the TAR.  Different TAR locations may be encompassed by different subTARs, in which case the functions to carry out this translation could be different. The extent of a subTAR is the region of the TAR for which its translation functions are valid. The definition of which subTAR is responsible for which TAR locations is given by the function $R{_A}$.

\begin{equation}
R{_A}: C_A \rightarrow S_A
\end{equation}

$R{_A}$ for a TAR $A$ maps every single TAR location to a subTAR in $S_A$ that encompasses it.

Moreover, the following relation holds true:

\begin{equation}
\forall s \in S_A \ \forall\ l_i \in \eta_s (R_A(l_i) = s)
\end{equation}

Every location in a subTAR $s$ extent is a location that maps to $s$ itself when applied to the subTARs location function $R{_A}$. As a consequence of this functional definition, we have the impossibility of a intersection of two subTARs extents for the same TAR. Any location in a TAR is associated to at most one subTAR. 

SubTARs have an associated position mapping function:

\begin{equation}
Fi_s: \mathbb{Z}^n \rightarrow \mathbb{Z}
\end{equation}

The function $Fi_s$ for a subTAR $s$ maps a n-dimensional TAR location to a single integer representing an index or offset for a data value into the linear storage scheme.
This linear address is then used to access the actual data values in the TAR. The subTAR data mapping function is responsible for this translation:

\begin{equation}
Fd_s: \mathbb{Z} \times D_A \rightarrow D 
\end{equation}

Where $D$ is the domain of a data element $e \in D_A$. The data mapping function maps a linear address along with a TAR data element to an atomic value in the domain of the respective data element given as the input for the function. Users may need to define data elements related to an entire TAR, or, more precisely, a data element that does not depend functionally on any dimension. This special type of data element is called a TAR property and $Fd_s$ becomes a constant.   

A TARS $\Gamma$ also includes the sets of roles $R_{\Gamma}$ and types $T_{\Gamma}$. A role is a string value that represents a special purpose data element with an important meaning in the application context. The set  $R_{\Gamma}$ contains all defined roles within a TARS. Roles are part of a type. A type $T$ is a triple $(N_T, M, O)$ where $N_T$ is a string containing the type's name, $M$ is a set of mandatory roles, and $O$ is a set of optional roles. Types allow users to give special meaning to every data element in a TAR. Special purpose operators in a TARS may take the TAR type into consideration. Some operators may only make sense for a TAR of a given type, since they depend on the existence of special purpose dimensions and attributes with well defined meaning in the application domain.

A TAR $A \in \Omega_{\Gamma}$ has an injective role mapping function defined as:

\begin{equation}
\Upsilon_{A} : D_A \rightarrow R_{\Gamma}
\end{equation}

The function $\Upsilon_{A}$ maps a data element to a role in the TARS, which indicates the role within the application context that the given data element fulfills. The notation $Type(A)$ refers to the type of the TAR $A$. A TAR $A$ is said to be of type $T$ if all data elements in $D_A$ are mapped to a role in $T_M$ or $T_O$, indicating the mandatory and optional roles of a type $T$, respectively. There must exist one element in $D_A$ that fulfills every mandatory role in $T_M$. Thus, we write: $Type(A) = T$  $\rightarrow$ $\forall \ d_e \in D_A$  $\ (\Upsilon_{A}(d_e) \in T_M) \lor$ $ (\Upsilon_{A}(d_e) \in T_O)$ $\land \quad \forall \ r \in T_M$  $\exists \ d_e \in D_A$ such that $\Upsilon_{A}(d_e) = r$. Two different data elements in $D_A$ cannot be mapped to the same role. This is guaranteed because $\Upsilon_{A}$ is an injective function. 

Relationships or links indicate that different data elements in the array schema correspond to the same entity. In a TARS, the $L_{\Gamma}$ is a set of relationships present in the schema. A relationship $R$ between TAR $A$ and TAR $B$ is a pair $(d_a, d_b)$ where $d_a \in A_{D_A}$ is a data element of $A$ and $d_b \in B_{D_B}$ is data element of $B$. Relationships are constraints that limit the domain of values that are valid in a data element given the current set of values held in another data element. This constraint can be expressed as:

\begin{equation}
{d_a}_{V_e} \subseteq {d_b}_{V_e}
\end{equation}

The set of every value held in the data element $d_a$ of $A$ is a subset of the data values held in the data element $d_b$ of $B$.

A TARS  has a set $\Theta_{\Gamma}$ of TAR operators that forms a TARS algebra. This algebra contains functions that allow users to create new TARs derived from the ones already defined. By combining the algebraic operators, users can express the most varied queries and analyses over data held in a one or more TARs. These operators can be type dependent, meaning that they work only with TARs of a given type. As stated before, a TAR of type T is guaranteed to have all mandatory roles of T, therefore an operators
that relies on a TAR A with type T assumes the values in A to be compliant with type T.

An operator $Op$ is defined as a triple  $(N_O, T_O, P_O)$ where $N_O$ is a string containing the operator name, $T_O$ is a list of types $<t_1, t_2, ... t_n> \  \in T_{\Gamma}$ defining the expected types for input TARS, and $P_O$ is a list $<p_1, p_2, ..., p_m>$ of atomic values forming the operator parameters set identifiers. 
New derived TARs can be expressed by using operators in a database query. A TAR $T_r$ resulting from the application of an operator $O_1$ on another TAR $T_o$ can be input into another operator $O_2$. Therefore, complex derived TARs can be created as the result of nested operator calls. For instance, a derived TAR $A$ can be produced by the query (where base\_tar is a TAR, and $v_a$, $v_b$, $v_c$ are parameters):  

\begin{equation}
A = O_1( O_2(\ ...\ O_n(base\_tar, v_n), \ ...\ v_b), v_a)
\end{equation}

This definition is the base for a functional query language, in which the functions are operations defined in TARS for a set of given types.

\subsubsection{Physical Specification}

In this section, we describe how the TARS data model is implemented in SAVIME. TARS structures are created in SAVIME with the use of the supported data definition language as it is usual for a DBMS.Users can define TARs, datasets, and types. Once a TAR is defined and a series of datasets is loaded into the system, it is possible to specify a subTAR by attaching datasets to it. 

\subsubsection*{Dataset}

A dataset is a collection of data values of the same basic type, like a column in a column-store DBMS. A dataset can contain data for a TAR attribute within a TAR region specified by a subTAR. The interpretation of datasets in SAVIME depends on the subTAR specification associated to it. If the data entries are ordered in some fashion, it is possible to take advantage of the preexisting ordering or structure of the data entries in the dataset. Otherwise, additional indexing information must be provided along the attribute values, since it cannot be inferred. Users are able to attach the same dataset into many different subTARS, which is a way to cope with the issue of partial functional dependencies with respect to dimensions mentioned previously.

\subsubsection*{Dimension Indexes}

TAR dimensions indexes form a domain of values that are represented in SAVIME in two main forms. It can be an implicitly defined range of equally spaced values. This is the simplest case in which all the user must specify is the lower and upper bounds of the range of values, and the spacing between two adjacent values. It is called implicit because these indexes do not need to be explicitly stored. For instance, the domain $D_i = (0.0,$ $,2.0$ $,4.0$ $,6.0$
$,8.0$ $,10.0)$ is defined by the lower bound $0.0$, the upper bound $10.0$ and all values are equally spaced in $2.0$ units. 

Dimensions whose indexes do not conform with these constraints have an explicit definition. In this case, the user provides a dataset specifying the dimension indexes. For instance, consider the following domain $D_e = (1.2,$ $,2.3$ $,4.7$ $,7.9$ $,13.2)$. It has a series of values that are not well-behaved and equally spaced, and thus, can not be represented implicitly. These types of dimensions are expected to emerge in regular grids with irregular granularities or even in the representation of sparse arrays.

Since these domains are a list of arbitrary numerical values (possibly non-integer), they require a mapping for an actual integer indexing scheme conforming to a traditional array representation. Therefore, SAVIME distinguishes between these non-integer index labels, called logical indexes, and the actual real integer indexes used as a storage reference by the system. The mapping between logical indexes depends whether the dimension is implicit or explicit. In the implicit case, it is straightforward to compute a zero based contiguous integer index from the range of values and vice-versa. In the explicit case, it is necessary to keep a dataset with this mapping and use the position of the logical index value within the dataset as the real integer index of the dimension.

\subsubsection*{Dimension Specification}

The actual data representation within the subTAR requires the combination between the dimension domain and the dimension specifications. All subTARs in a TAR have a list of dimension specifications, one for each dimension in the TAR.  These dimension specifications define the TAR region the subTAR encompasses, but they also are a fundamental part in the implementation of the mapping functions. These functions are defined conceptually in the model, but are implemented considering six possible configurations between dimension specifications types and dimension types.

A dimension specification can be of three different types. It can be ORDERED, meaning that the underlying datasets are completed filled within the specification range regarding a dimension. A specification can be PARTIAL, meaning that only a subset of the dimension values are filled for a given dimension, but this subset remains coherent throughout other dimensions, and also can be TOTAL, which is the least efficient but more general representation. In the TOTAL representation, all indexes, for all dimensions need to be stored explicitly, since the TAR region is sparsely populated and the underlying datasets contain only data entries for actual non-empty cells.

\begin{figure}[ht]
	\centering
	\includegraphics[width=3.3in]{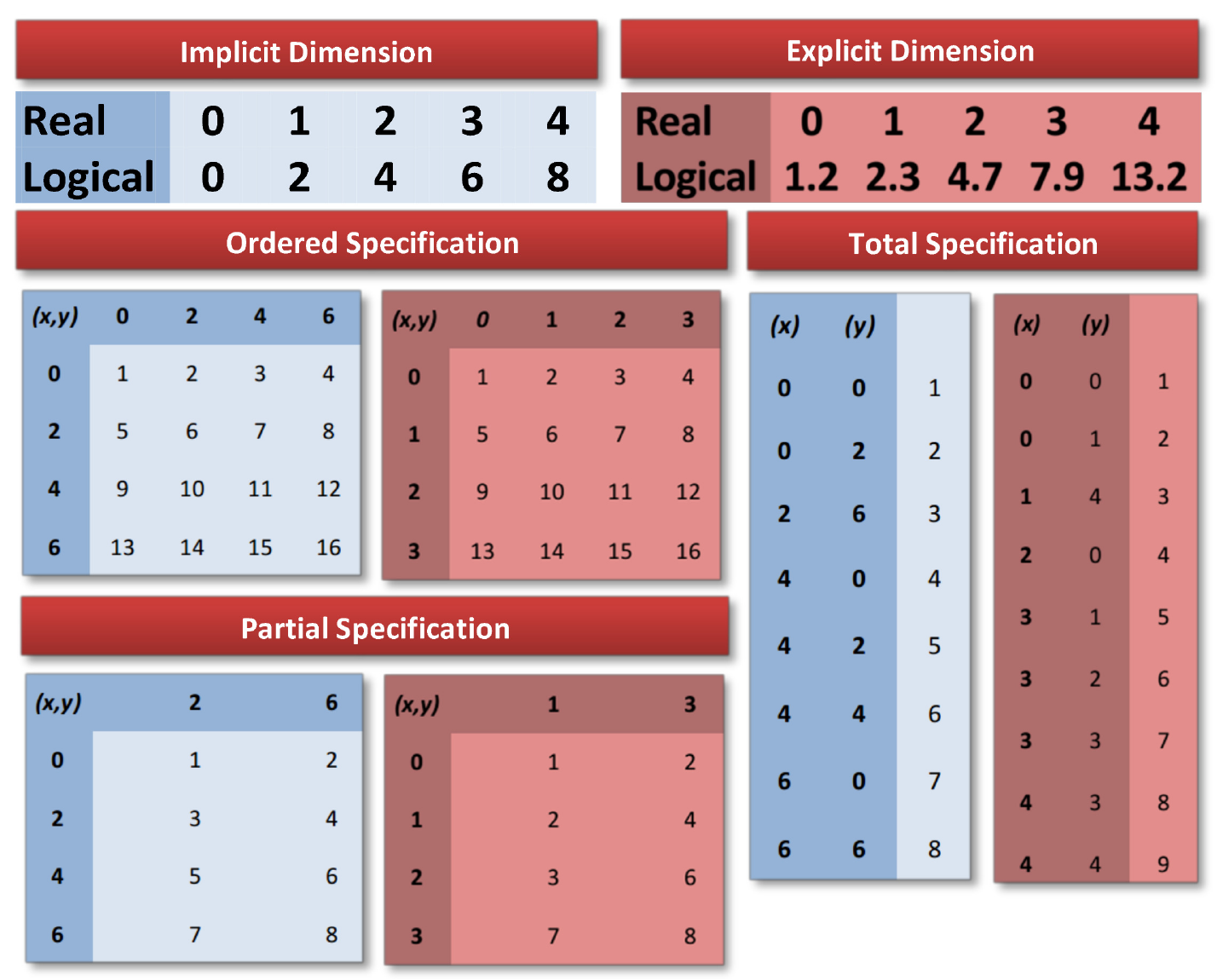}
	\caption{Possible implementations for a subTAR in SAVIME.}
	\label{fig:dimtypes}
\end{figure}

Combining the two possible dimension types, with these three possible dimension specification types, we end up with six possible configurations as shown in Figure \ref{fig:dimtypes}. Consider an example TAR with two dimensions $x$ and $y$ and with a single numerical attribute in its cells.  This TAR can be implemented by having implicit or explicit dimensions. At the top of Figure \ref{fig:dimtypes}, there are the mapping for an implicit (blue) and explicit (red) dimensions. The boxes represent possible subTARS definitions, the blue ones represent subTARs for an example TAR whose dimensions are implicit, and the red ones are subTARs for an example TAR whose dimensions are explicit. The values for the TAR attribute within the subTARs region is given by a dataset filled with a list of ordered integer values from $0$ to $N$. Values at each cell illustrates their positioning in the array according to the SubTAR memory layout specification.

The ordered specification illustrated in Figure \ref{fig:dimtypes} is the simplest case in which all possible cells are filled with some values (non-sparse). In the example, specifications are given in a row-major order. This order is decided by the user during the definition of the subTAR and  must be in accordance with the order of the values within the dataset for the attributes. Regardless of the dimension specification type, the ranges for the region a subTAR  encompasses is given by the real indexes that defined a lower and upper bounds on each dimension. 

Nevertheless, we see that the dimension indexes along the data values differ from the implicit and explicit examples. The dimension indexes stored along a subTAR are different depending on the type of the dimension. For implicit dimensions, logical indexes are stored and for explicit dimension, real indexes that need to be mapped into logical ones are stored. The ordered dimension specification is the most efficient one, because it does not require neither logical or real indexes to be stored. All dimension indexes for any cell can be inferred by the cell position within the dataset. Also, slice and dice operations over a subTAR with this configuration are easy, due to the fact that all the data is sorted in a known fashion.

Another possibility is a partial specification. In this case, we have a fully defined $x$ dimension, and a partially defined $y$ dimension. Partial dimension specifications allows the representation of sparse data, in a very specific case in which the sparsity happens in a regular pattern. In the example, we see that only positions $2$ and $6$ in the implicit example (or positions $1$ and $3$ that are equivalent to logical indexes $2.3$ and $4.7$ respectively) are filled with data. However, these positions are filled across the entire dimension $x$ also, therefore, we need to store explicitly only the indexes for the filled positions and we can take advantage of the fact that these indexes occur in the same pattern for the whole subTAR region. These indexes are kept in a dataset that is as large as the length of the filled dimensions. In this case, since there are only two positions filled in the $y$ dimension, we store a dataset with size 2. The logical values are stored in the case of implicit dimension, and a series of real indexes are stored in the case of an explicit dimension.

Total specification is the most general possible representation of a subTAR and the least efficient one. Only non-sparse subTARs have ordered specifications, and only subTARs with a very specific kind of sparsity can be represented by partial specifications. But any array, regardless of how sparse it is, can be represented by this scheme. Total specifications require all dimension values (either real or logical) to be explicitly stored. In practice, it is like a degenerated TAR representation that resembles simple tabular data. Datasets with fully materialized dimension indexes must be kept.Thus, if a subTAR for a bi-dimensional TAR has a length $m$, two datasets with length $m$ need to be stored containing the index of every dimension for every cell in the TAR. In this scenario, there is not only the overhead of storing indexes, but also the least efficient execution of slice and dice operations. Data has no longer a well-behaved structured, therefore SAVIME needs to carry a full subTAR scan in order to filter out data.

While ordered and partial specifications can coexist for different dimensions in a subTAR, the same is not true for total dimension specifications. Total representation refers to all dimension specifications at once, meaning that it is impossible to have an ordered and a total dimension specification in the same subTAR. On the other hand, each subTAR is independent from the perspective of its dimension specifications. As long as two subTARs do not intersect with each other (which is forbidden by the TARS data model), each one of the subTARs can have its own dimension specification and dimension ordering. Thus, a subTAR can represent a very dense TAR region, while another one can represent a very sparse one, and they can all coexist and hold data for the same TAR.

\subsection{SAVIME Architecture} \label{arc}

\begin{figure}[ht]
	\centering
	\includegraphics[width=3.7in]{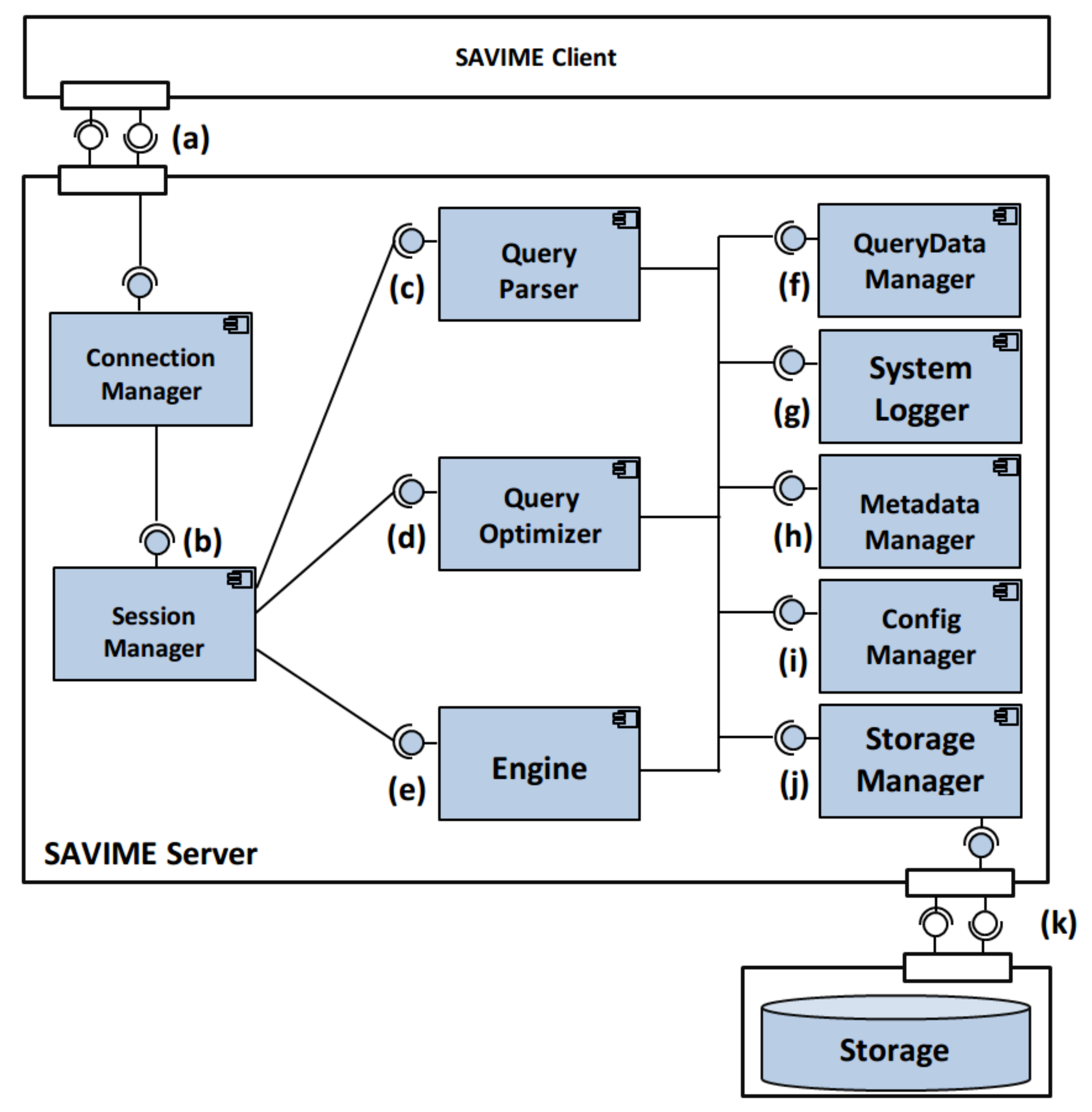}
	\caption{SAVIME components diagram}
	\label{fig:arc}
\end{figure}

SAVIME has a component-based architecture common to most DBMSs. It was built considering a modular pattern, as shown in Figure \ref{fig:arc}, with a series of basic components that encapsulate its main capabilities.  

A SAVIME client communicates with the SAVIME server by using a simple protocol that allows both ends to exchange messages, queries and datasets. When a notification reaches the server, the first module to be activated is the connection manager, which in turns triggers the session manager. The session manager fires a new thread to answer the user's requests. This thread calls the parser, responsible for turning a query submitted by the client in his request into a query plan. The query plan is then passed on to the optimizer and finally to the execution engine. Once the engine has enough data, it notifies the session manager, which sends query results back to the client. 

The query parser, the query optimizer and the engine use the services provided by other modules, such as the metadata manager, the storage manager, the system logger and the configuration manager. The metadata manager does the bookkeeping of the schema components defined by the user. The system logger and configuration manager serve respectively to keep the activities log and give access to a key value store of configuration values. The query data manager is a component whose functions is to hold important structures related to a query being executed, i.e., the actual query text, the query plan, references for datasets being exchanged, and some other bookkeeping information.

The storage manager is a critical component responsible for creating, removing, accessing and operating over datasets on disk or in memory. SAVIME maintains datasets as files, preferably in a memory based file system. The storage manager uses memory-mapped files, allowing SAVIME to access the datasets as a memory buffer. If files are kept in a memory based file system, the usage of memory-mapped files makes the kernel memory where the data is stored accessible in SAVIME's address space. Thus, no copies between the kernel and local buffers are necessary during query processing and data ingestion. The major drawback of memory-mapped files is the occurrence of many minor page faults that can affect performance negatively \cite{Tevanian87aunix}. To alleviate the effects of page faults, temporary datasets, created during query processing, can be kept in a huge pages file system, that uses huge memory pages. This approach minimizes the total number of pages needed and accessed by SAVIME, which, consequently, reduces the number of page faults in orders of magnitude.

Figure \ref{fig:arc} indicates the main interfaces between components. They are:

\begin{itemize}
 \item (a) Interface between the client and the server, based on the SAVIME protocol and the supported query language. SAVIME can also be deployed with a library called libstaging, that allows for the efficient harvest of data in the client address space and the transfer and loading into SAVIME.
 \item (b) Interface between the connection manager and the session manager. The connection manager notifies the session manager about new requests and messages, and the session manager uses the connection manager's interface to send responses back to the client.
 \item (c) A query parsing interface, which receives a query and returns a query plan. 
 \item (d) An interface for creating optimized query plans from the plans produced by the parser. 
 \item (e) Interface for executing the query plan.  
 \item (f) Interface for temporarily storing query information, such as the query string, the query plan and incoming datasets. 
 \item (g) Interface for logging system activities. 
 \item (h) Interface for creating and removing TARs, subTARs, types, datasets and other structures. 
 \item (i) Interface for accessing a key-value configuration cache.
 \item (j) Interface for creating, removing, copying, filtering and many more operations over datasets.  
 \item (k) The storage manager accesses some underlying storage where the datasets are kept, more commonly the storage is some file system interface provided by the operating system.
\end{itemize}

All modules are currently implemented as a series of C++ classes, each one of them with an abstract class interface and an underlying concrete implementation. Ideally, adding support to different query languages, storage schemes and protocols should be possible by modifying only the correspondent module.

\subsubsection{Query Processing} \label{engine}

SAVIME supports a functional query language which was designed in the spirit of \cite{marathe1999query}. A SAVIME query is a text containing a series of operators defined as functions. Operators can be nested in order to form a more complex query. Most of them expect one or more input TARs, and originate a newly created output TAR. Unless a special operator is called to materialize the query final TAR, it is generated as a stream of subTARs, sent to the client and then discarded. 

\begin{figure}[ht]
	\centering
	\includegraphics[width=3.5in]{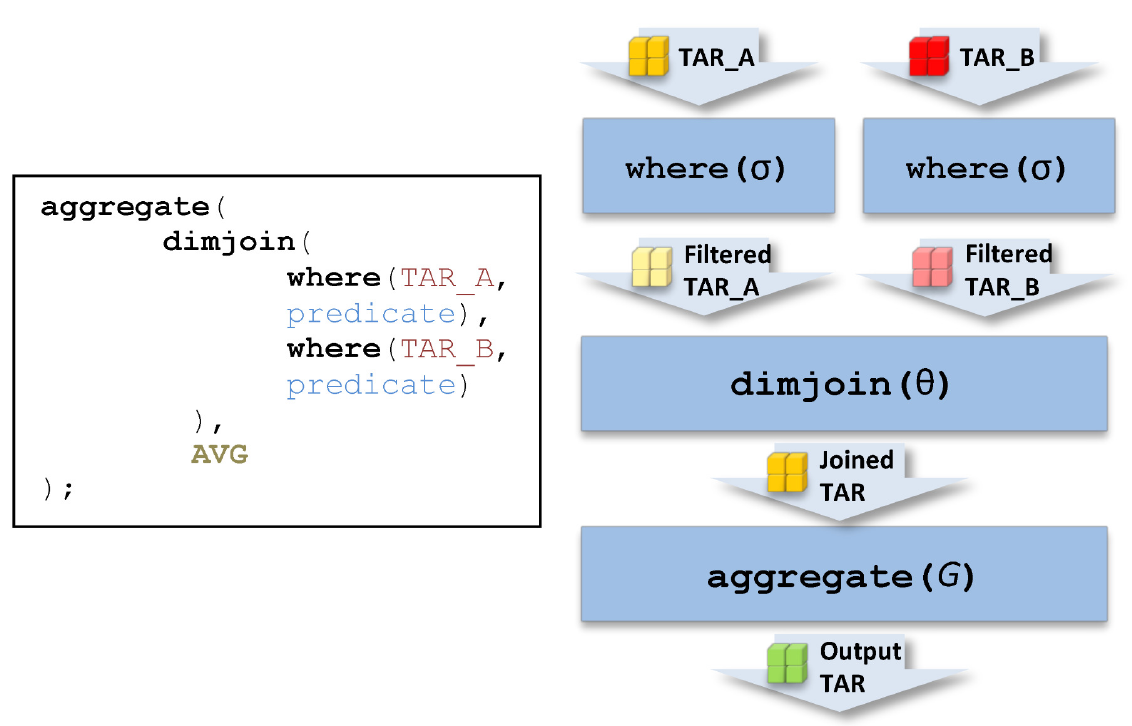}
	\caption{A query in SAVIME functional query language.}
	\label{fig:query}
\end{figure}

Figure \ref{fig:query} contains an example of query in SAVIME. The operator WHERE, that filters out TAR data according to a predicate, is nested with the DIMJOIN operator, that joins two TARs based on their index values, and the final TAR is created as the output of an aggregation function calculated by the AGGREGATE operator.

Since SAVIME is designed to be a tool to help analysis and visualization of simulation data, its operators are expected to provide means for executing the most varied types of analysis. A list of the main operators currently supported by SAVIME is presented in Table \ref{tb:op}.

\begin{table}[ht]
\centering
\caption{SAVIME main operators.}
\small
\label{tb:op}
\def\arraystretch{1.1}%
\begin{tabular}{|l|l|} \hline
\textbf{OPERATOR} & \textbf{DESCRIPTION}\\ \hline
\shortstack[l]{SELECT \\ $\ $} & \shortstack[l]{$\ $\\ Projects dimensions and attributes. \\ $\ $}\\ \hline
\shortstack[l]{WHERE \\ $\ $} & \shortstack[l]{$\ $\\ Filters data according to a complex logical\\ predicate considering dims and attributes.  }\\ \hline
\shortstack[l]{SUBSET \\ $\ $} & \shortstack[l]{$\ $\\ Creates an n-dimensional data slice\\ according to specified bounds.  }\\ \hline
\shortstack[l]{DERIVE \\ $\ $} & \shortstack[l]{$\ $\\ Adds an attribute with derived values\\ calculated from a user defined expression.  }\\ \hline
\shortstack[l]{CROSS\_JOIN \\ $\ $} & \shortstack[l]{$\ $\\ Creates the cartesian product between\\ cells in two TARs.  }\\ \hline
\shortstack[l]{DIM\_JOIN \\ $\ $} & \shortstack[l]{$\ $\\ Equivalent to an equijoin considering \\  pairs of matching dimension indexes.  }\\ \hline
\shortstack[l]{AGGREGATE \\ $\ $} & \shortstack[l]{$\ $\\ Summarizes data by evaluating \\  common aggregation functions.  }\\ \hline
\shortstack[l]{CATALYZE \\ $\ $ \\ $\ $} & \shortstack[l]{$\ $\\ Converts a series of TARs representing \\  mesh data to a viz file and optionally runs a \\ catalyst script with it.  }\\
\hline\end{tabular}
\end{table}

\subsubsection*{Array Operation Semantics}

Operators can either keep, add or remove data elements, i.e., dimensions or attributes from a TAR. Due to this fact, there is a general rule for TAR typing during query execution. Each TAR operator outputs a TAR whose type is either the type of one of the input TARs or none. For instance, if a TAR undergoes a projection operation in which only a subset of its data elements is present, either all mandatory roles for the original type are fulfilled and the output TAR keeps the type, or some mandatory role is dropped, and the output TAR is then typeless. The same is true for aggregation operations. For operation such as filters that do not change the input TAR, there is simply no action regarding the type, it is just replicated in the output TAR. For operations that combine two TARs, such as the cross product and the joins, the type of one of the input TARs is maintained in the output TAR. Therefore, depending on the query executed, the final output TAR might conserve the type of the TARs used to generate it. It is very important in queries that rely on types, such as the visualization queries that will be discussed in the next section. 

\subsubsection*{Execution Model}

From the perspective of the execution model, SAVIME operates TARs as a subTAR stream pipelined across operators. It is a compromise between an execution model that operates a data tuple or data entry at a time, like the Volcano execution model \cite{Graefe:1994:VEP:627290.627558}, and an execution model like in MonetDB, in which entire intermediate tables are materialized at each step of query processing \cite{Boncz2005MonetDBX100HQ}. The first execution model is not adequate for in memory processing of large datasets, and the second might not be feasible for complex queries with very large intermediate TARs, which would need to be fully materialized in memory before passing on to the next operator.

Therefore, each operator in SAVIME works on a single subTAR at a time and is implemented as an independent function or module. In most operators, SubTARs are processed serially, one after another. The execution of an operator on a subTAR, however, requires a lot of tasks over large datasets. These tasks are parallelized with OpenMP \cite{openmp} constructs. It is an assumption in SAVIME that subTARs are big enough to justify the creation of threads to process them individually in parallel.

Operators in a query have some input TARs and generate a single output TAR. Figure \ref{fig:query} illustrates these operators. The parser, during the process of creating the query plan, defines the schema of every single output TAR for every operation. The first operators in the query have as input TARs already stored in SAVIME. The final output TAR is the one sent to the client application which fired the query. Each operator knows exactly the schema for its input TAR, and also the form of the output TAR to be generated. However, the number and characteristics of every single subTAR can only be known at run time. Thus, a subTAR stream flows through the operators. Each operator implementations gets subTARs from the previous operator and provides subTARs for the following operator in the query plan. There is not a one-to-one relationship between input and output subTARS, since some operators might need to read many or even all subTARs from its input TARs before outputting even a single subTAR. It is the case, for instance, of the AGGREGATE operator, which must process all the input subTARs in order to create a correct output for the following operator.

During query execution, each operation consumes the subTARs generated by previous operations as it generates subTARs to be consumed by following operations. Every operator must know which input subTARs are going to be needed in order to generate a given output subTAR when requested by the next operation in the pipeline. 

When a subTAR for an intermediate TAR is generated and passed on to the next operator, it is maintained in a temporary subTARs cache. These subTARs contain their own group of datasets that naturally require a lot of storage. Therefore, once a subTAR is no longer required by any operator, it must be removed from memory, since it is very likely that a complex query, with many intermediate TARs would rapidly consume all available memory during query execution otherwise. An operator implementation is agnostic regarding its surroundings, and does not know when to free or not a subTAR. All the operators need to know is when it will not require a given subTAR any longer. When this happens, the operator notifies the execution engine that it is done with a given subTAR.  

However, since the same subTAR can potentially be input into more than one operator during a query, freeing it upfront might not be a good idea, because it might be required again. In this case, SAVIME would have to recreate it. To solve this problem, every subTAR has an associated counter initially set to the number of operators that have its TAR as their input. When an operator notifies the engine that it no longer needs that specific subTAR, the respective counter is decreased. Once the counter reaches zero, all operators possibly interested in the subTAR are done, and now it is safe to free the subTAR. 

This approach always frees the used memory as soon as possible and never requires a subTAR to be created twice. However, some operators, like DIMJOIN and AGGREGATE might require many subTARs to be kept in memory before freeing them. In an environment with limited memory, it would not be feasible to cope with very large subTARs in this case. A solution then, would be the adoption of a more economical approach, trading off space with time by freeing and regenerating the subTARs whenever memory is running low. The problem of deciding which datasets to free during query execution, and when to free them, is a research problem in the context of in-memory big data analytics \cite{7584962} and it is not the focus of this paper.

\section{Visualization Use Case} \label{viz}

Simulation is the process of designing a computational model of a system to understand and predict its behavior \cite{dym2004principles}. Simulations are particularly useful in situations where it is hard or even impossible to execute real tests to acquire data. They depend on the creation of mathematical models describing the relation between physical quantities. 

A mathematical model captures the behavior of a phenomenon, with equations, usually solved by numerical methods. In many cases, the selected numerical method relies on the solution of a system of linear equations. Therefore, the actual solution is a numerical array kept in the solver's memory space. 

In addition, some methods require the discretization of the domain in a form of a grid or mesh. Depending on the domain, modelers can adopt either structured or unstructured meshes, divided into cells or elements. Meshes have topological and geometrical representations. Geometrical aspects are related to shapes, sizes and absolute positions of their elements, such as points. Topology representation captures the relations between elements, like their neighborhoods or adjacency, without considering their position in time and space. The actual data output from the simulation comprises what is called field data, with scalar and vector quantities defined over every grid point.

In transient simulations, these field data values vary not only in space but also in time. Furthermore, researchers and modelers need to run their simulation many times with different parameters and compare the output from different runs. Thus, a complete simulation dataset provides data values varying through space, time and different trials, along with the complete domain specification of its mesh (topology and geometry) and the parameters used for every specific simulation run. 

Visualization is one of the main activities that scientists execute with data generated by simulations. In this context, viz tools and libraries, such as VTK and Paraview Catalyst \cite{Ayachit:2015:PCE:2828612.2828624} are widely used to allow researchers to gain insights about scientific datasets. 

\begin{figure}[ht]
	\centering
	\includegraphics[width=3.3in]{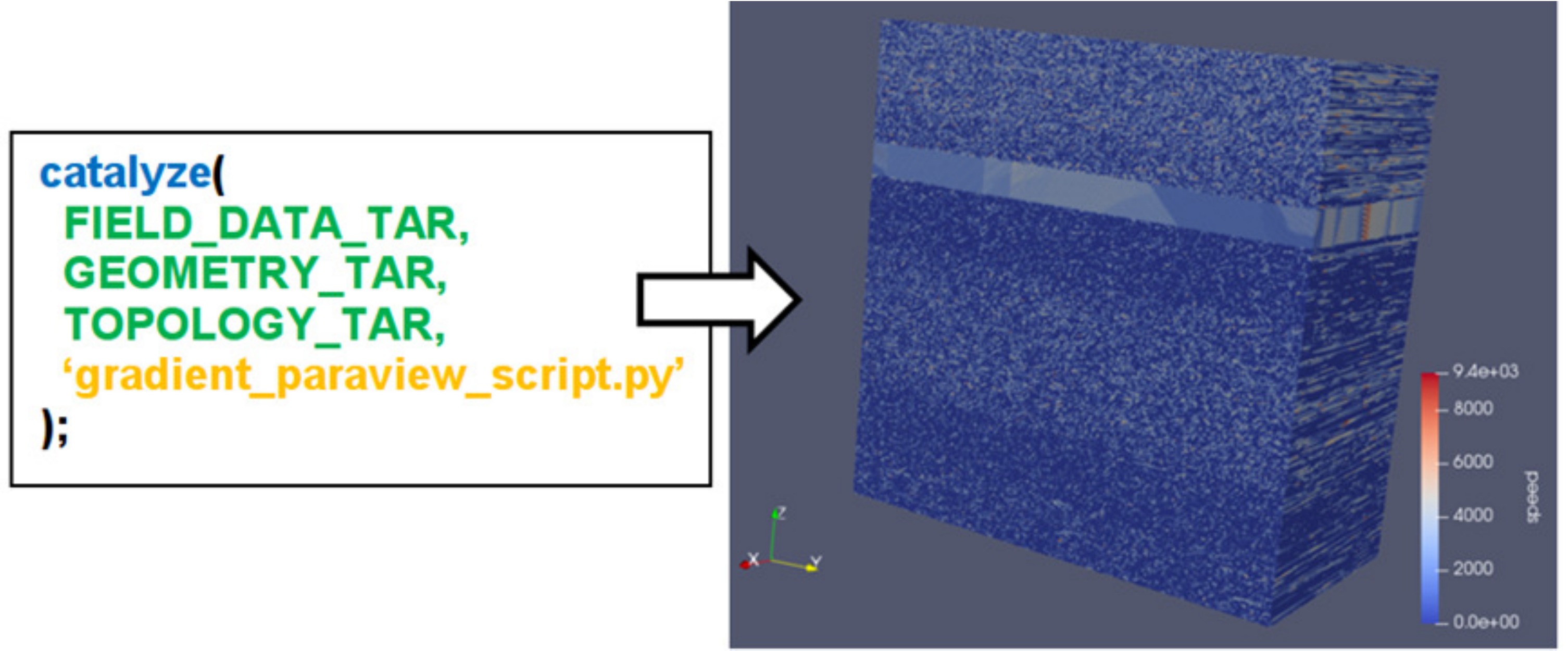}
	\caption{Visualization query using Catalyze operator.}
	\label{fig:viz}
\end{figure}

SAVIME is able to generate VTK data files to be directed imported by Paraview, or even to execute Catalyst analytical scripts. Visualization in SAVIME is done with the use of a special purpose operator named CATALYZE, shown in Figure \ref{fig:viz}, which gets the resulting TARs of a query, converts it to the VTK structures and optionally executes a Catalyst script with the query output.

\begin{figure}[ht]
	\centering
	\includegraphics[width=5.3in]{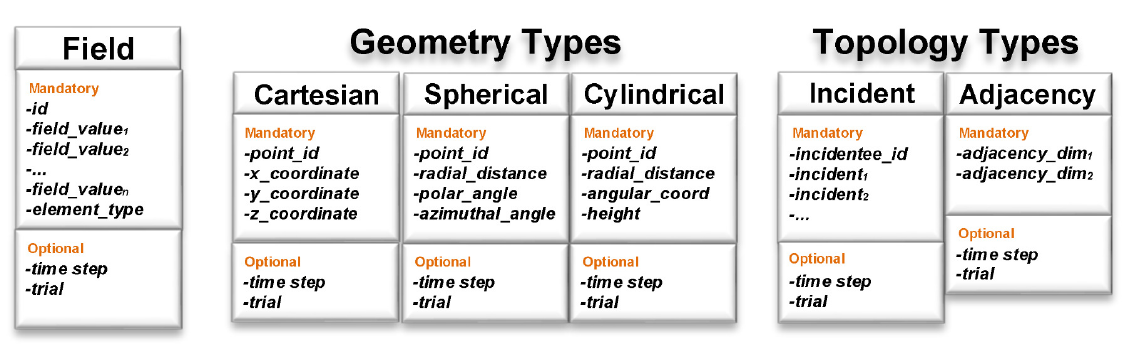}
	\caption{TAR types for data visualization.}
	\label{fig:types}
\end{figure}

In order to implement this operator and visualize data, SAVIME needs to be aware of which parts of the dataset represent the geometry and the topology of the mesh, and how the field data is laid out in this schema and then create the respective VTK structures. The TARS data model enables this semantic representation by defining special TAR types for topologies, geometries and field data (Figure \ref{fig:types}). For instance, a Cartesian Geometry type has a mandatory role id for identifying points, and other roles for x, y and z coordinates. There is also the same optional roles time step and trial. A mesh geometry can change or evolve during the same trial, in a scenario in which the application domain is deformable, or a different mesh can be used for every simulation run.

Data related to the mesh topology requires special types, such as the Incident Topology and Adjacency Topology. The Incident Topology type captures the semantics of topologies specified as a series of incident relationships and the Adjacency topology type captures topologies represented as an adjacency matrix. Since the mesh can be different in every time step for the same run, or for different runs of the same model, the optional roles time step and trial are also present.

Since Catalyst is embedded in SAVIME, it is possible to create visualization files and even run scripts with data harvested directly from SAVIME's memory space. Without such support, users would need to move data out of SAVIME's memory space and convert it before being able to visualize it, which is very inefficient. 

\section{Experimental Evaluation} \label{exp}

The author conducted experimental evaluations in order to compare SAVIME, SciDB (version 16.9) and a third approach based on the usage of NetCDF files (version 4.0), used as a baseline. The goal of this evaluation is to demonstrate how data ingestion in SAVIME is more efficient than in the current state of the art array systems.

\begin{figure}[ht]
	\centering
	\includegraphics[width=5.3in]{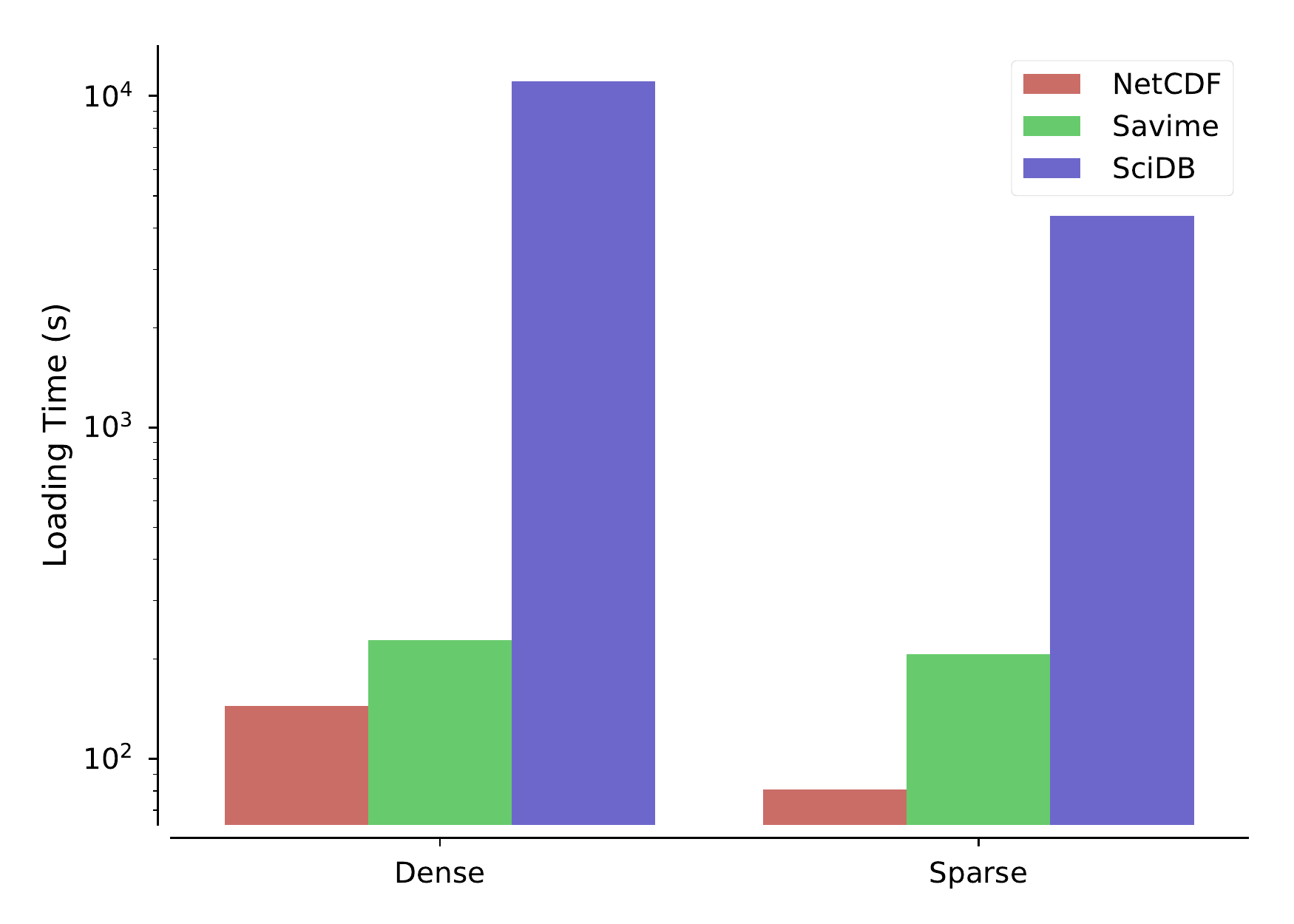}
	\caption{Ingestion time Experiment Results}
	\label{fig:exp1}
\end{figure}

As we can see in Figure \ref{fig:exp1}, the ingestion time taken by SciDB is almost 20 times larger than the time taken by SAVIME, due to costly rearrangements needed on data to make it conform with the underlying storage configuration. Besides, there is an extra overhead during the lightweight data compression done by SciDB, which makes the dataset roughly 50\% smaller when stored but increases loading time prohibitively. We load data into SciDB using the most efficient possible methods supported, combining the REDIMENSION and INSERT operators as advised in the official community forums \cite{paradigm4forum}. 

The author also evaluated the performance considering the following types of queries (with dense and sparse multidimensional arrays):

\begin{itemize}
 \item Window query consists of retrieving a subset of the array defined by a range in all its dimensions. The performance for this query depends on how data is chunked and laid out. High selectivity queries, which need to retrieve only a very small portion of the data tends to be faster than full scans. Therefore, we compared low and high selectivity queries, filtering from a single to all 500 tiles.

 \item Exact Window Query is the easiest possible query for an array database. It consists of retrieving data for a single tile, meaning the system has close to zero work filtering out the result.
\end{itemize}

\begin{figure}[ht]
	\centering
	\includegraphics[width=5.3in]{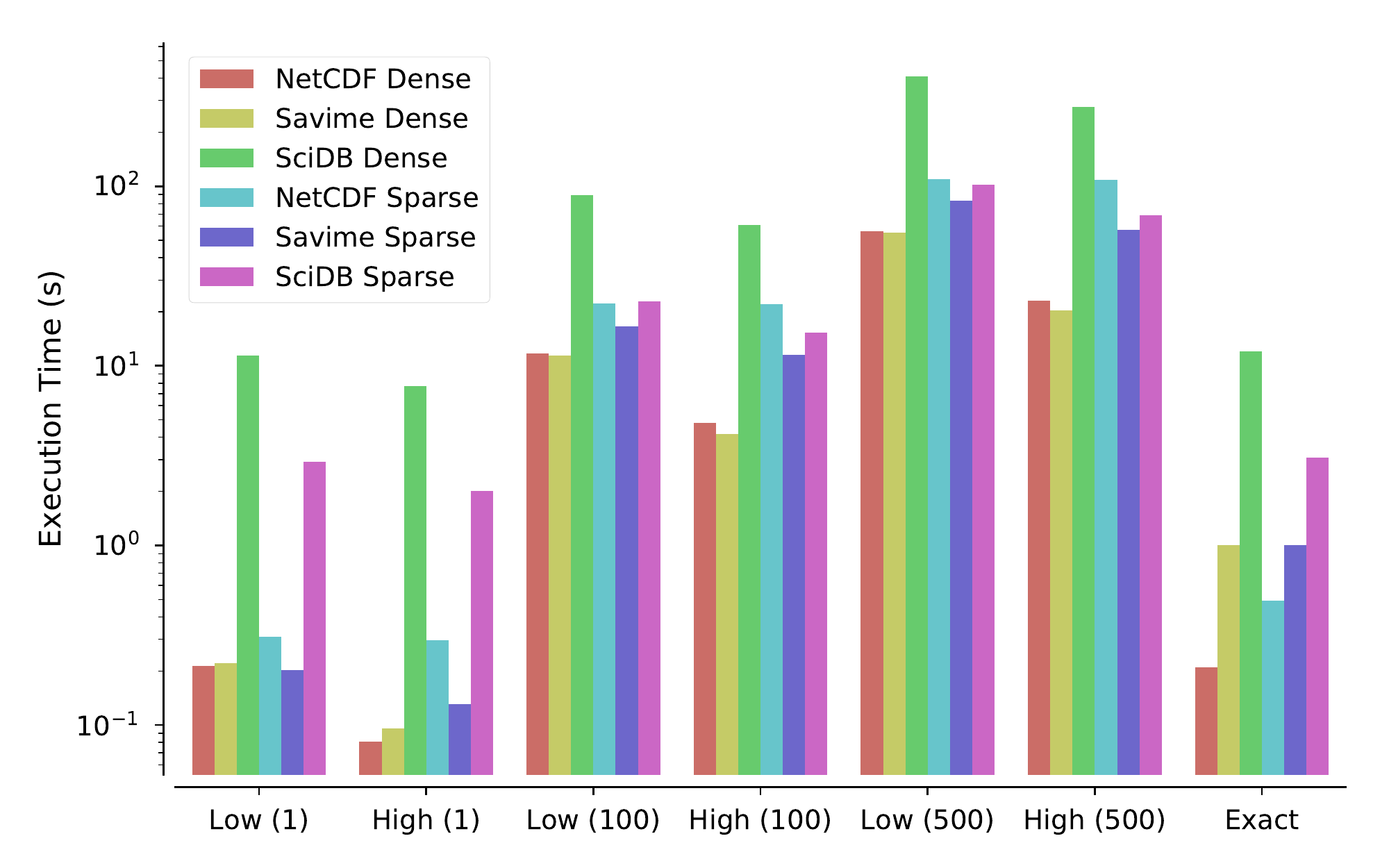}
	\caption{Query Execution for the Dense Dataset}
	\label{fig:exp2}
\end{figure}

The experimental results are shown in Figure \ref{fig:exp2}. The average time of all runs are presented along with the corresponding intervals of confidence. We considered window queries with low selectivity (over 70 \% of all cells in a tile) and high selectivity (around 20 \% of all cells in a tile), and intersecting with only one, a hundred or even the total five hundred tiles. 

It is noticeable that SAVIME either outperforms SciDB or is as efficient as it in all scenarios. The most important aspect of this result is that, even without any previous data preprocessing, SAVIME is able to simply take advantage of the existing data structure to answer the queries efficiently, which validates the model as a feasible alternative to existing implementations.

The results show that, for any storage alternative in both dense and sparse formats, the subsetting of a single tile is very efficient. The differences shown for the exact window query and for low and high selectivity window queries that touch a single tile are very close. SciDB takes a few seconds in most cases, while SAVIME takes in average 1 second. NetCDF is the most efficient in this scenario, retrieving desired data in less than a second.

However, for queries touching 100 or 500 chunks, we can see the differences between querying dense and the sparse arrays. The dense dataset is filtered more efficiently, since it is possible to determine the exact position of every cell and read only the data of interest. It is not possible for sparse data, since one is not able to infer cell positions within the tiles. In this case, every single cell within all tiles that intersect with the range query must be checked.

In dense arrays, we can observe a reduced time for retrieving data in high selectivity queries in comparison with low selectivity queries. The execution time of window queries should depend only on the amount of data of interest, since cells can be accessed directly and thus, no extra cells need to be checked. The execution times considering 100 or 500 tiles in SAVIME and NetCDF are in accordance with this premise. However, SciDB shows poorer performance, being up to 8 times slower. It is very likely that SciDB needs to process cells outside of the window of interest depending on the compression technique and the storage layout adopted. SciDB seems to be more sensible to tiling granularity, requiring fine-grained tiles that match the window query to have a performance similar to the NetCDF approach. 

There is not much to be done for querying sparse arrays except for going through every cell in the tiles intersecting the window specified by the query. This can be seen in Figure \ref{fig:exp2} as all alternatives show very similar performance. The main difference is that for achieving this result with NetCDF, an OpenMP application needed to be written, while the same result could be obtained with a one-line query in SAVIME and SciDB.

We should consider, of course, that this evaluation is designed to show the ideal scenario for SAVIME, which is a shared memory multi-core fatnode with a considerable amount of memory. SciDB is a natively distributed system, not necessary optimized for a single-node in memory processing. Moreover, SAVIME makes use of memory-mapped files in a hugetblfs, i.e., a special in memory file system with huge memory pages that increases performance of applications that rely on memory-mapped files. We choose this environment, because it is interesting for an in-transit simulation data post-processing workload, since data does not need to be stored on disk before being analyzed. Nevertheless, the results show that even state-of-the-art systems, in their current form, cannot cope well with data from simulation applications in an in-transit approach, and thus, a specialized solution can perform better. 

\subsection{Integration with Numerical Solver} \label{sav_v_mhm}

This section presents the evaluation of the amount of overhead imposed to the simulation code when integrating with SAVIME. We use the simulation tools based on the MHM numerical method \cite{DBLP:journals/corr/GomesPVP17} as a representative numerical simulation application. We compare three approaches. In the first approach, SAVIME is used IN-TRANSIT, in a single node (fatnode) while the simulation code runs in a different set of nodes, and thus data needs to be transferred. In the second approach, SAVIME is used IN-SITU, with individual SAVIME instances running on each  node, the same used by the simulation code. In this scenario, the data does not need to be transferred, since it is maintained in a local SAVIME instance that shares the same computational resources used by the simulation code. In the third approach SAVIME is not used, but instead, the data is stored in ENSIGHT files (the standard file format used by MHM), and analysis are performed by an ad-hoc Message Passing Interface application in Python. This last scenario serves as a baseline implementation, thus we call it the baseline approach. The computational resource used is  the Petrus cluster  at DEXLab, with 8 nodes, each with 96 GB of RAM and 2 Intel(R) Xeon(R) CPU E5-2690 processors. 

To evaluate the integration between SAVIME and a simulation tool based on the MHM solver, we use a 2D transport problem over a mesh with 1.9 million points. We run the simulation up to the 100th time step, and store either in SAVIME (approaches 1 and 2) or in an ENSIGHT file (approach 3), data from 50\% of all the computed time steps. In both cases, data is always kept in a memory based file system, and never stored on disk. Once data has been generated, it is analyzed with a PARAVIEW pipeline that carries out the computation of the gradient of the displacement field of the solution. This part is either done by a special operator in SAVIME, or by an AD-HOC MPI Python application using the Catalyst library (baseline), depending on the approach being run. Additionally, we measure the cost of running the simulation without any further analysis, to highlight the \textit{simulation only} cost.

\begin{figure}[ht]
	\centering
	\includegraphics[width=5.9in]{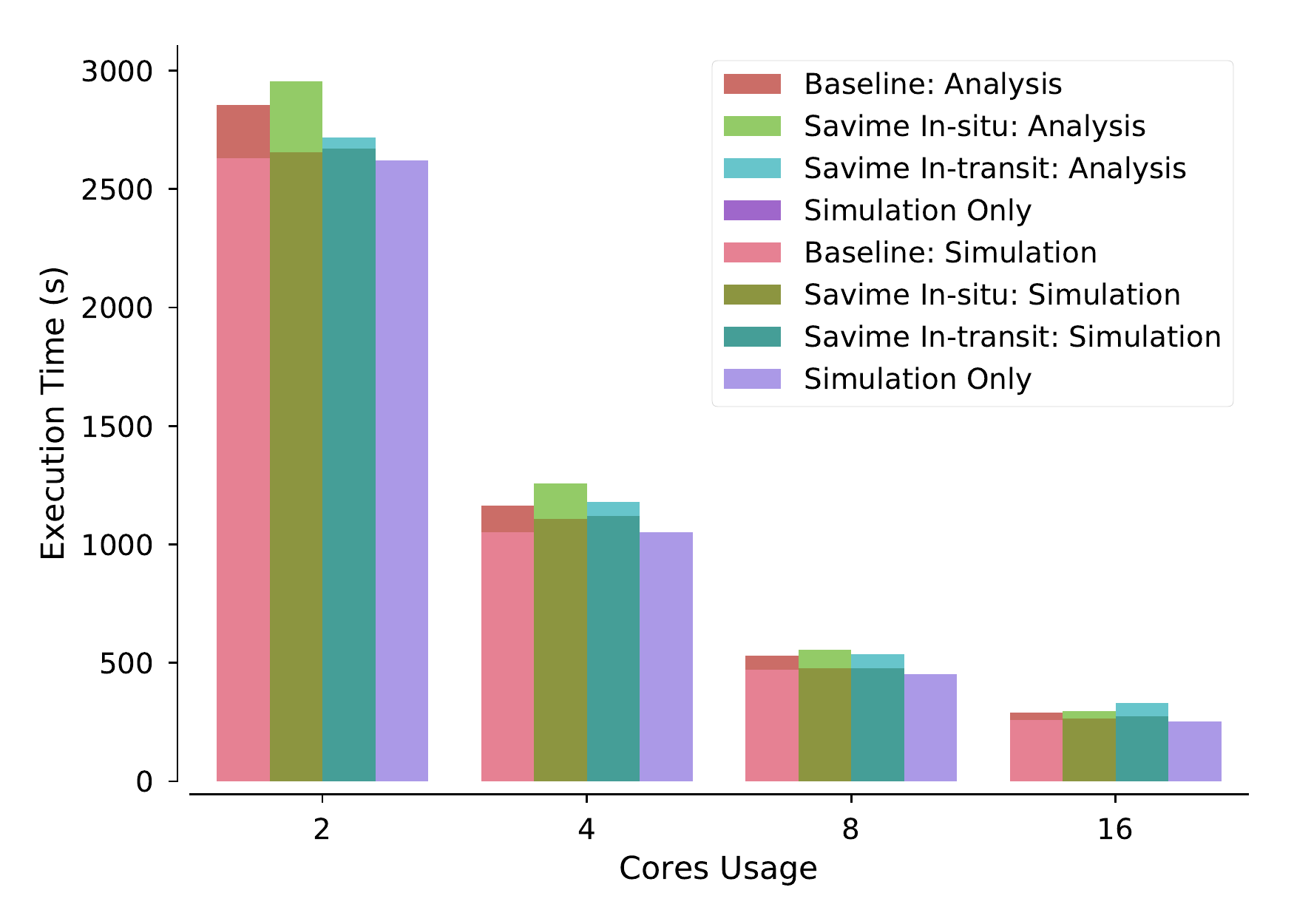}
	\caption{Simulation and Analysis Scalability}
	\label{fig:exp22}
\end{figure}

Figure \ref{fig:exp22} shows the results when running the three approaches, varying the amount of MPI processes spawned or the number of cores used by the MHM simulation code. In this experiment, the simulation code runs and produces its results and then, the simulation output data is read and processed in the analysis step. The plot shows, for each evaluated number of MPI processes, the simulation time and the analysis time as stacked bars. The graph shows that the cost of the analysis process is significantly smaller than the cost for computing the simulation.  Moreover, as the \textit{Simulation Only} run shows, the overhead introduced by storing the data in SAVIME or as an ENSIGHT file is negligible, which confirms the claim that SAVIME can be introduced into the simulation process without incurring in extra overhead. From the point of view of the effect of SAVIME on simulation scalability, the storage of data in SAVIME does not impair the capability of the simulation code to scale up to 16 cores.

The IN-TRANSIT approach differs from the other two approaches since it uses a separate computational resource to execute the analysis step. As we see in Figure \ref{fig:exp22}, even when we increase the number of cores the simulation code uses, the analysis time does not change, because the analysis step is done in the fatnode, and always uses the same number of cores (16) independently from the actual number of cores used by the simulation code. The IN-TRANSIT approach illustrates a scenario in which all data is sent to a single computational node and kept in a single SAVIME instance. This approach offers some extra overhead and contention, since all data is sent to a single SAVIME instance, but this enables posterior analysis that transverse the entire dataset without requiring further data transfers.

The SAVIME IN-SITU approach uses the same computational resources used by the simulation code. When we increase the number of cores used by the simulation code, we also increase the numbers of cores used by SAVIME for analysis. The same is true for the baseline approach, meaning that the AD-HOC application also uses the same number of cores the simulation code uses. Even though the SAVIME IN-SITU approach is slightly slower than the baseline approach, we see that both are able to scale similarly. The difference observed in performance between using SAVIME and coding a specialized application becomes less significant as we increase the number of cores being used during the analysis phase. Nevertheless, the small performance loss in this case might be justified by the convenience of using a query language to express analysis instead of the extensive and error prone process of coding other applications to execute analysis.

\section{Conclusion}  
The adoption of scientific file formats and I/O libraries rather than DBMSs for scientific data storage is due to the impedance mismatch problem that increases data ingestion time, and also due to the overhead of moving data out of the database for the creation of visualizations.

To mitigate these problems, and also offer the benefits of a DBMS for simulation applications, the author proposes in its PhD thesis, the research and development of a system called SAVIME. 

This work showed how SAVIME, by implementing the TARS data model, does not impose the huge overhead present in current database solutions for data ingestion, while also being able to take advantage of preexisting data layouts to answer queries efficiently. Our preliminary results show how SAVIME compares to SciDB, and how it can be plugged into the simulation application pipeline. 

Future work will focus on array processing optimization techniques; the implementation of a comprehensive set of array operators, and the evaluation of the feasibility of applying SAVIME to manage other data than simulation data, such as machine learning models and training datasets.

\bibliographystyle{unsrt}  
\bibliography{references}  

\end{document}